%
%
\magnification=\magstep1
\baselineskip=11pt plus .1pt minus .1pt
\hsize=12.5truecm
\vsize=19.0truecm  
\hfuzz=5pt\vfuzz=5pt
\tolerance=1000
\overfullrule=0pt
\parskip=0pt
\abovedisplayskip=3 mm plus6pt minus 4pt
\belowdisplayskip=3 mm plus6pt minus 4pt
\abovedisplayshortskip=0mm plus6pt minus 2pt
\belowdisplayshortskip=2 mm plus4pt minus 4pt
\predisplaypenalty=0
\clubpenalty=10000
\widowpenalty=10000
\parindent=2em
%
%
\font\pgnumfont=cmr9
\font\headlinefont=cmti9
 \font\titlefont=cmbx10
\font\authorfont=cmr10
\font\addressfont=cmti9
\font\datefont=cmr9
\font\sumfont=cmr9
\font\itl=cmti9

\font\absfont=cmbx9
\font\secfont=cmr10
\font\subsecfont=cmti10
\font\subsubsecfont=cmr10
\font\figfont=cmr9
\font\figheadfont=cmbx9

\font\tabheadfont=cmbx9
\font\mainfont=cmr10
\font\petitrm=cmr9

%
%
%
\newtoks\TITLE \newtoks\AUTHOR \newtoks\ADDRESS \newtoks\SUMMARY
\newdimen\sumindent \sumindent=\parindent
\newtoks\KEYWORDS \newtoks\SUBMITTED \newtoks\ACCEPTED
\newtoks\SENDOFF
%

%
%
\newtoks\firstpage
\let\firstpage=Y
\newtoks\AUTHORHEAD \newtoks\ARTHEAD \newtoks\VOLUME \newtoks\PAGES
\if!\the\AUTHORHEAD!\AUTHORHEAD={\the\AUTHOR}\fi
\if!\the\ARTHEAD!\ARTHEAD={\the\TITLE}\fi
\footline={\hfil}
\headline={\ifodd\pageno\rightheadline \else\leftheadline\fi}
\def\leftheadline{\if Y\firstpage\firsthead\global\let\firstpage=N
  \else\lefthead\fi}
\def\rightheadline{\if Y\firstpage\firsthead\global\let\firstpage=N
  \else\righthead\fi}
\def\lefthead{\pgnumfont\number\pageno\hfil\headlinefont\the\AUTHORHEAD}
\def\righthead{\headlinefont\the\ARTHEAD\hfil\pgnumfont\number\pageno}
\def\firsthead{\headlinefont Baltic Astronomy,~vol.\the\VOLUME,
\the\PAGES,~\the\year .\hfil}
\voffset=2\baselineskip 
%

\newdimen\oldbaselineskip \oldbaselineskip=\baselineskip
\def\test#1{\newlinechar=`@\if!\the#1! \message{#1 not given@}\fi}%
\def\printheader{
  \parindent=0pt
  \null\vskip1.cm
  \test{\TITLE}
  \vbox{\baselineskip=15pt
    \titlefont\the\TITLE
    }
  \vskip8mm plus8mm
  \test{\AUTHOR}
  \authorfont\the\AUTHOR
  \vskip2mm
  \test{\ADDRESS}
  \addressfont\the\ADDRESS
  \vskip2mm
  \test{\SUBMITTED}
  \line{\datefont Received \the\SUBMITTED
    \if!\the\ACCEPTED!\else, accepted \the\ACCEPTED\fi.\hfill}
  \vskip4mm plus4mm
  \vbox{\leftskip=\sumindent\parindent=0pt
    \parskip=5pt
    \absfont Abstract.
    \test{\SUMMARY}
    \sumfont\the\SUMMARY\par
    \absfont Key words:
    \test{\KEYWORDS}
    \sumfont\the\KEYWORDS\par
    }
  \sumfont
  \if!\the\SENDOFF!\else\footnote{}{Send offprint requests to:
 \the\SENDOFF}\fi
  \parindent=2em
  }
%
%
\newdimen\uppergap \newdimen\lowergap
\uppergap=5mm \lowergap=3mm
\newdimen\secind \newdimen\subsecind \newdimen\subsubsecind
\setbox0=\hbox{\secfont 9. }\secind=\wd0
\setbox0=\hbox{\subsecfont 9.9. }\subsecind=\wd0
\setbox0=\hbox{\subsubsecfont 9.9.9. }\subsubsecind=\wd0
\def\section#1{\goodbreak\par\vskip\uppergap
  \noindent\hangindent\secind\hangafter=1\secfont#1
  \vskip\lowergap\mainfont\par\nobreak}
\def\subsection#1{\goodbreak\par\vskip\uppergap
  \noindent\hangindent\subsecind\hangafter=1\subsecfont#1
  \vskip\lowergap\mainfont\par\nobreak}
\def\subsubsection#1{\goodbreak\par\vskip\uppergap
  \noindent\hangindent\subsubsecind\hangafter=1\subsubsecfont#1
  \vskip\lowergap\mainfont\par\nobreak}
%
%
\def\WFigure#1#2#3{\goodbreak\midinsert\vbox{
  \null\centerline{#2}\vskip5truemm
  \figheadfont\indent Fig.~#1.\figfont\ #3
  \par\mainfont
  }\endinsert}
%

%

%

%

%
\newdimen\tabind
\setbox0=\hbox{\tabheadfont Table 55.} \tabind=\wd0

%
%
\def\References{\vskip\uppergap
\line{\secfont REFERENCES\hfill}
  \vskip0.8\lowergap
 \petitrm
  }
\def\ref{\goodbreak
\hangindent12pt\hangafter=1
\noindent\ignorespaces}
\def\endref{\egroup}
%
%
\def\byebye{\egroup\par\vfill\supereject\end}
%
%

%
%

\def\degr{\hbox{$^\circ$}}

\def\arcmin{\hbox{$^\prime$}}
\def\arcsec{\hbox{$^{\prime\prime}$}}
\def\utw{\smash{\rlap{\lower5pt\hbox{$\sim$}}}}
\def\udtw{\smash{\rlap{\lower6pt\hbox{$\approx$}}}}




\def\ddown{\lower2.5ex\hbox}
\def\ddow{\lower1.7ex\hbox}
\def\down{\lower1ex\hbox}
\def\uppp{\raise1ex\hbox}
\def\dnnn{\lower1ex\hbox}
\def\uuppp{\raise2ex\hbox}

\def\ts{\thinspace}
\def\(o-c){$O-C$}


\def\angstr{A\kern-.56em\raise1.9ex\hbox{$\scriptscriptstyle\circ$}$\,$}

\newdimen\free\newdimen\shift
\def\Entry#1#2#3{\par\goodbreak\smallskip%
  \setbox1=\vbox{\advance\hsize by-10mm\parindent=0pt
    \def\\{\par}%
    \it#1. \rm#2}
  \line{\box1\hfill#3}\smallskip
}%
\newdimen\savesize

\def\shiftfigure #1#2#3#4#5{
    \vbox to #2 { \ifodd #5 \rightskip#4 \else\leftskip#4 \fi
                  \null\vfil
                  \figheadfont Fig.~#1.\figfont #3
                  \medskip
                }
                          }

\year2002

\input psfig.sty
\def\ts{\thinspace}

\year 2002
\VOLUME {~11}
\PAGES {205--218}
\pageno=205

\TITLE={CCD PHOTOMETRY AND CLASSIFICATION OF STARS IN THE NORTH AMERICA
AND PELICAN NEBULAE REGION. I. MOL\.ETAI PHOTOMETRY}

\AUTHOR={V. Laugalys and V. Strai\v zys}

\AUTHORHEAD={V. Laugalys, V. Strai\v zys}
\ARTHEAD={North America and Pelican nebulae region. I}

\ADDRESS={Institute of Theoretical Physics and Astronomy,
Go\v stauto 12, \hfil\break Vilnius 2600, Lithuania;
vygandas@itpa.lt, straizys@itpa.lt}

\SUBMITTED={April 10, 2002}

\SUMMARY={Magnitudes and color indices in the {\itl Vilnius} seven-color
system are measured for 690 stars down to $\sim$13.2 mag in the area of
the North America and Pelican nebulae.  Spectral types, absolute
magnitudes, color excesses, interstellar extinctions and distances of
the stars are determined.  The plots of interstellar extinction {\itl
A}$_V$ versus distance for the North America Nebula and for the dark
cloud L935 show that both areas are covered by the same absorbing cloud,
situated at a distance of 600 pc.  The maximal extinction in the
area of the nebula is $\sim$3 mag, while in the dark cloud L935 it is
much greater.}

\KEYWORDS={stars:  fundamental parameters, classification, multicolor
photometry, Vilnius photometric system -- ISM:  dust, extinction,
individual objects:  cloud L935}

\printheader

\section{1. INTRODUCTION}

This work continues the investigation of the area containing the North
America and Pelican nebulae complex, using photometry of stars in the
Vilnius seven-color photometric system and their two-dimensional
classification.  In the first paper (Strai\v zys, Mei\v stas, Vansevi\v
cius \& Goldberg 1989a, Paper I) the results of photoelectric photometry
of 249 stars down to 11 mag, their two-dimensional classification, color
excesses, interstellar extinctions and distances have been given.  In
the second paper (Strai\v zys, Goldberg, Mei\v stas \& Vansevi\v cius
1989b, Paper II) interstellar extinction in the area was investigated.
The dark cloud L935, separating the North America and Pelican nebulae,
was found to be at a distance of 550 pc.  In the third paper (Strai\v
zys, Kazlauskas, Vansevi\v cius \& \v Cernis 1993, Paper III) the cloud
distance was revised to 580 pc, using photometry and classification of
additional 564 stars down to 12.5 mag.  In the above papers, the scale
of absolute magnitudes of stars was based on the Hyades distance modulus
$V$--$M_V$ = 3.2.  If we accept the new distance modulus of 3.3
(Perryman et al. 1998), the cloud distance determined in the last paper
changes to 610 pc.  \vskip0.5mm

The present paper starts a new series of investigations of the North
America and Pelican nebulae complex, using CCD photometry in the {\it
Vilnius} system obtained with the 35/51 cm Maksutov telescope of the
Mol\.etai Observatory in Lithuania and with the 1 meter Ritchey
telescope at the US Naval Observatory at Flagstaff, Arizona. The present
paper is based on the Mol\.etai observations only.

\section{2. OBSERVATIONS}

CCD frames were obtained in 2000 and 2002 with the 35/51 cm Maksutov
telescope at the Mol\.etai Observatory in Lithuania.  The camera, loaned
from the Tr{\o}mso University Observatory (Norway), was fixed in the
Newtonian focus, having the scale of 2.87$\arcmin$/mm or
4.25$\arcsec$/pixel.  The field size was 25$\times$25 mm$^2$ or
1.2$\times$1.2 sq. degrees.  The camera has a thinned back-illuminated
Tektronix 1012$\times$1012 pixel chip with 25 $\mu$m pixels and with a
thermoelectric cooling down to --40$\degr$C.  \vskip0.5mm

The focal sizes of images are very small:  even at a seeing of
3$\arcsec$ they are 17 $\mu$m only, i.e., smaller than the CCD pixel
size.  Therefore, the exposures were made with a small defocusing.
Usually a star image contained from 4 to 9 pixels.  A set of round (60
mm diameter) filters of the {\it Vilnius} system was used.  The
ultraviolet filters $U$ and $P$ are glass filters and $X$, $Y$, $Z$, $V$
and $S$ are interference filters.  The exposure times in different
filters were from 0.5 to 4 min.  Each night in each filter several
twilight flats were obtained.  The centers of the exposed areas were
shifted aiming to cover the 2$\times$2 sq. degree field bounded by the
coordinates (2000.0):  RA = 20$^{\rm h}$51.5$^{\rm m}$ -- 21$^{\rm
h}$00$^{\rm m}$, DEC = +43$\degr$30$\arcmin$ -- +45$\degr$30$\arcmin$.
A schematic map of the investigated area is given in Figure 1 with the
contours of the dark cloud corresponding to $A_V$ = 5 mag according to
Cambr\'esy et al.  (2002).

Additionally, we have measured magnitudes and color indices
of 150 stars in the {\it Vilnius} system photoelectrically. The
observations were done with the 165 cm telescope of the Mol\.etai
Observatory in 1999.

\WFigure{1}{\psfig{figure=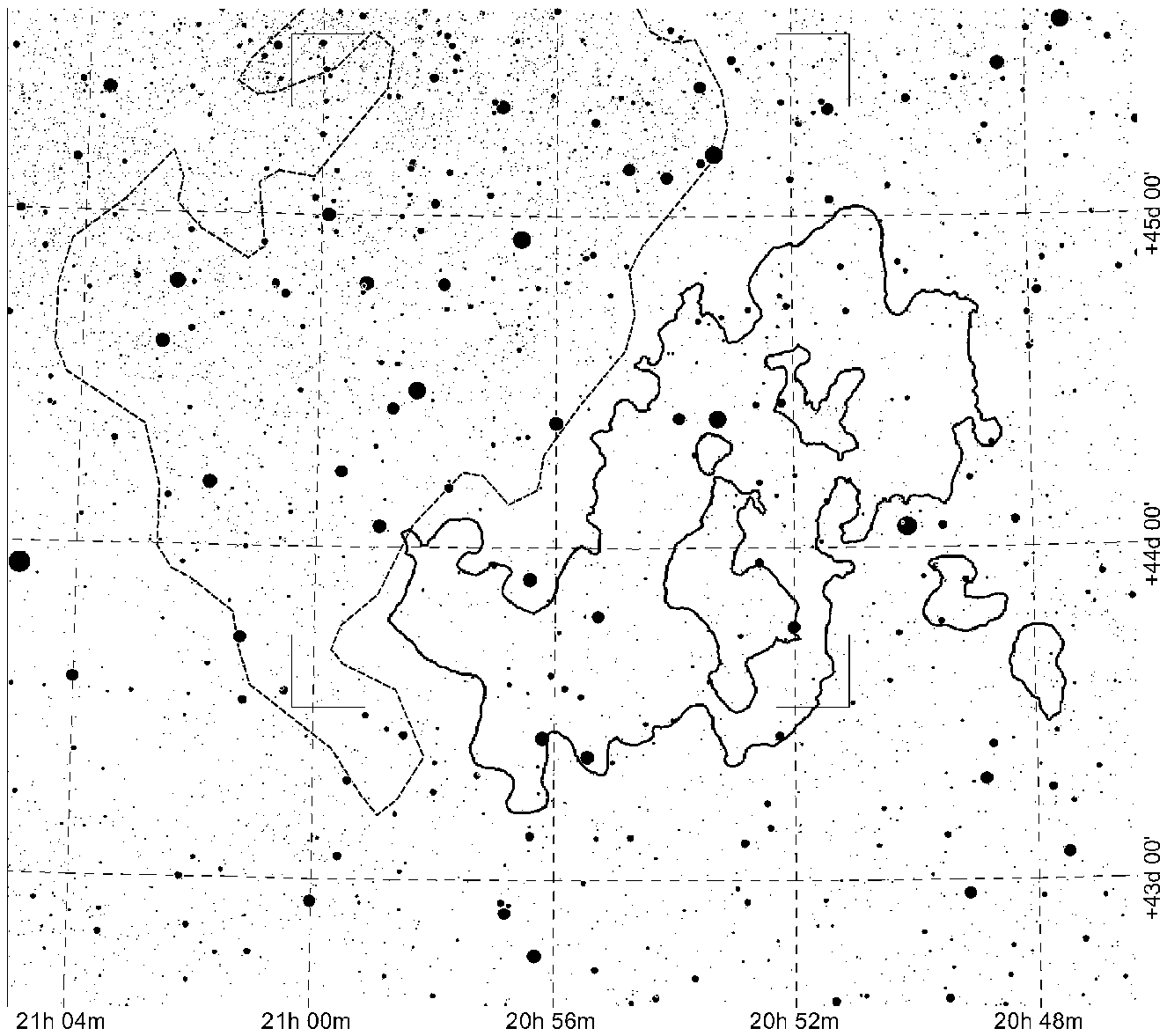,width=12.4truecm,angle=0,clip=}} {
Schematic map of the investigated area and its surroundings.  The angles
mark the corners of the investigated area.  The broken line shows the
contours of the North America Nebula.  The solid line shows the contours
of the dark cloud corresponding to {\itl A}$_V$ = 5 mag according to
Cambr\'esy et al.  (2002).  The grid of the coordinates is for 2000.0.}

\section{3. REDUCTIONS}

For obtaining the instrumental magnitudes of stars the multi-aperture
method of the IRAF program package was used.  The size of the aperture
used was 15--20$\arcsec$.  Instrumental $V$ magnitudes and color indices
were transformed to the standard {\it Vilnius} system by color equations
obtained by comparing about 60--100 standard stars in the same field
observed photoelectrically and taken mostly from Strai\v zys et al.
(1993).

Transformation equations obtained from observations were verified by
synthetic photometry, calculating the magnitude differences between the
standard and the CCD systems by the equation:
$$
m_{\rm {st}}\!-\!m_{\rm {CCD}}~=~-2.5\log {\int {F(\lambda)R_{\rm
{st}}(\lambda) \tau^x(\lambda)d\lambda}\over \int{F(\lambda)R_{\rm
{CCD}}(\lambda) \tau^x(\lambda)d\lambda}} + const, \eqno(1)
$$
where $m_{\rm {st}}$ and $m_{\rm {CCD}}$ are magnitudes defined by the
response functions $R_{\rm {st}}(\lambda)$ and $R_{\rm {CCD}}(\lambda)$,
$F(\lambda)$ is the energy flux distribution function in the spectrum of
a star, $\tau^x(\lambda)$ is the transmittance function of interstellar
dust of $x$ unit masses (interstellar extinction law).  The energy
distribution curves for 25 stars of various spectral classes and
luminosity classes V--IV--III were taken from Strai\v zys \&
Sviderskien\.e (1972) with the corrected ultraviolet, as described by
Strai\v zys et al.  (1996).  The response curves of the standard Vilnius
system were taken from the Strai\v zys (1992) monograph, Table 59.  The
response curves of the CCD system were obtained by multiplying the
sensitivity function $s\ts (\lambda)$ of the CCD chip, the transmittance
functions of the filters $f(\lambda)$ and the meniscus lens of the
telescope $m\ts (\lambda)$, and the reflection functions of two
aluminized mirrors $a^2(\lambda)$.  For the ultraviolet filter the mean
atmospheric transmittance function $p\ts (\lambda)$ at zenith was taken
into account.  The functions are taken from the following sources:
$s\ts (\lambda)$ -- from the manufacturer's description of the Tektronix
1024$\times$1024 CCD camera, $f(\lambda)$ -- from the measurements with
a photoelectric spectrometer, $m\ts (\lambda)$, $a\ts (\lambda)$ and
$p\ts (\lambda)$ -- from Strai\v zys (1983).  \vskip0.5mm

The coefficients of synthetic color-equations for indices $P$--$V$,
$Y$--$V$, $Z$--$V$ and $V$--$S$ are close to those of the equations
determined from observations.  For color indices $U$--$V$ and $X$--$V$
some nonlinearity of the synthetic equations was found.  The response
curve of the instrumental magnitude $U$ is shifted toward long
wavelengths by the meniscus lens.  This lens is made from ultraviolet
transmitting glass; however it is $\sim$30 mm thick and this makes its
transmittance in the ultraviolet wavelengths $<$330 nm rather low.  This
causes the mentioned nonlinearity in the transformation equation of
$U$--$V$.  In case of $X$--$V$, the nonlinearity originates from the
shift of the transmittance curve of the filter onto the H$\delta$ line.
\vskip0.5mm

The accuracy of the final magnitudes and color indices is seen from
Figure 2 which shows the instrumental errors given by the IRAF package.

\WFigure{2}{\psfig{figure=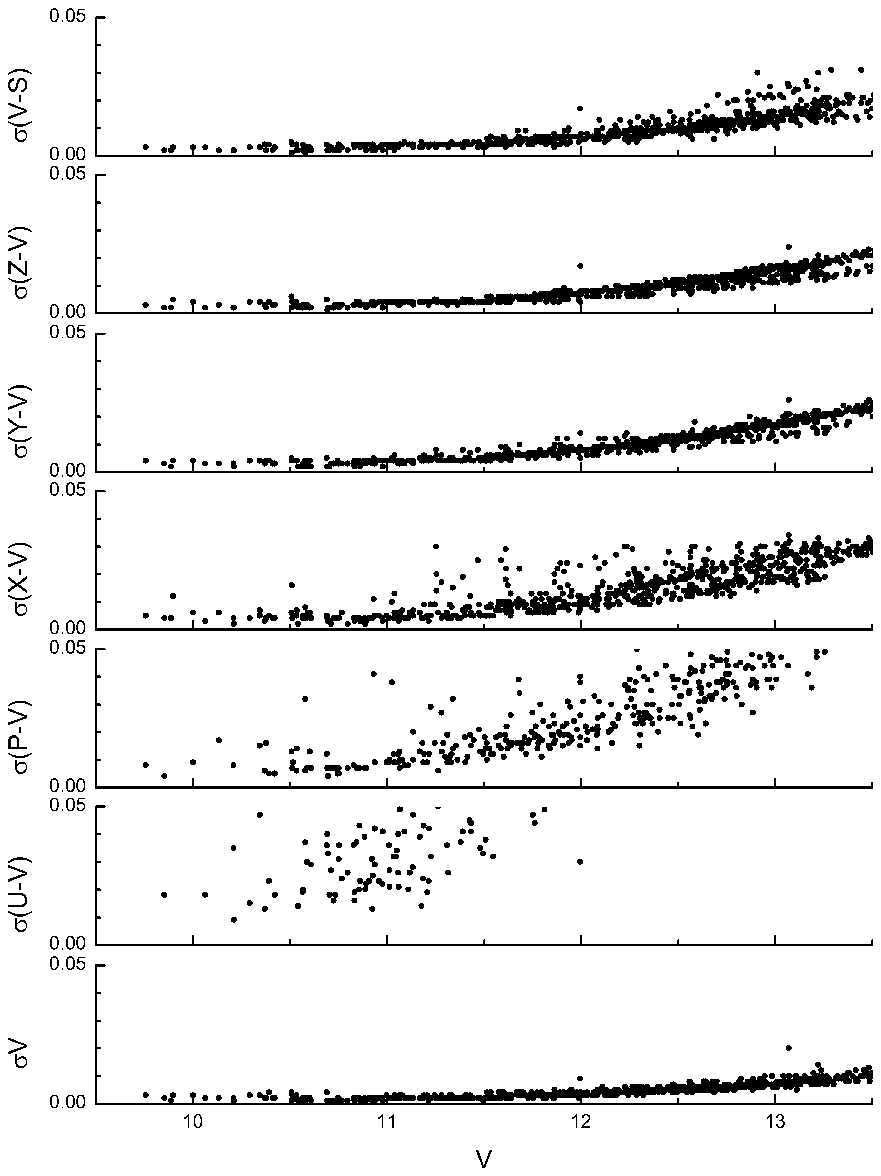,width=8truecm,angle=0,clip=}} {
Instrumental errors of magnitudes  and color
indices as a function of the magnitude {\itl V}. The ``sequences"
of dots seen in the graph are due to slightly different exposure times
and background intensities on different frames.}

The stars which had double or multiple images in the CCD frames were
omitted from photometry.  However, the binary stars (both physical and
optical) with a separation of $<5\arcsec$ are unresolvable in our CCD
images and seem as single stars.  Trying to find more binaries, all
stars down to the limiting magnitude (13.2 mag in $V$) were verified for
duplicity on the Internet's virtual telescope SkyView of NASA based on
the DDS (Digital Sky Survey) scans of the Palomar atlas red and blue
plates (http://skyview.gsfc.nasa.gov).  About 80 stars showing double or
multiple images or close neighbour stars were also rejected from further
analysis.

\section{4. THE CATALOG AND CLASSIFICATION OF STARS}

The final catalog will be published elsewhere.  It contains 690 stars
down to $V$ = 13.2 mag observed by CCD and 150 stars down to $V$ = 12.0
mag observed photoelectrically, 130 stars are common.  For 27\% of stars
all color indices are available, 53\% of stars are without $U$--$V$ and
20\% -- without both $U$--$V$ and $P$--$V$.

     For the stars with all six color indices available,
spectral types were determined by using the method of matching of 14
different interstellar reddening-free $Q$-parameters of a program star
to those of about 8300 standard stars of various spectral and luminosity
classes, metallicities and peculiarity types from the General
Photometric Catalog of Stars Observed in the Vilnius System (Strai\v zys
\& Kaz\-laus\-kas 1993). The $Q$-parameters are defined by the equation:
$$
Q_{1234} = (m_1-m_2) - (E_{12}/E_{34})(m_3-m_4), \eqno(2)
$$
where $m$ are the magnitudes in four (sometimes three) passbands,
$m_1-m_2$ and $m_3-m_4$ are the two color indices and $E_{12}$ and
$E_{34}$ are the corresponding color excesses.  The $E_{12}/E_{34}$
ratio slightly depends on spectral type, and this dependence is taken
into account by iterations.  The ratios are taken for the Cygnus
interstellar reddening law which, according to the study of Strai\v zys,
Corbally \& Laugalys (1999), is valid in the North America and
Pelican nebulae area.
\vskip0.5mm

The matching of $Q$-parameters leads to a selection of some standard
stars with a set of $Q$s most similar to those of the program star.  The
match quality is characterized by
$$
\sigma Q = \pm\sqrt{{\sum_{n}^{} \Delta Q_i^2}\over n}, \eqno (3)
$$
where $\Delta Q$ are differences of corresponding $Q$-parameters of the
program star and the standard, $n$ is a number of the compared
$Q$-parameters (in our case, $n$ = 14).  For the stars observed with an
accuracy of $\pm$0.01 mag, the $\sigma Q$ value is of the order of
$\pm$0.01--0.02 mag.  In such a case the match is considered to be
sufficiently good, and the spectral type (spectral class + luminosity
class) of the standard star may be prescribed to the program star.  In
our case, for the program star we have accepted the average spectral and
luminosity classes of the three to five best matching stars.  Since the
errors of the observed color indices for a part of the stars is $>$0.01
mag, their classification accuracy is lower.  If the matching accuracy
was of the order of $\pm$0.03 mag or larger, the star was not classified
at all. There are 30 such stars ($\sim$4\%) in the catalog. Some of them
may be unresolved binaries or peculiar stars. The known binary stars
with $\rho\leq$15$\arcsec$ and $\Delta V\leq$3 mag were also not
classified.

The accepted spectral types are in the table available from the authors
in electronic form.  Table also contains spectral types found in other
publications, mostly in Schwassmann \& van Rhijn (1938), Metik (1960),
Kharadse et al.  (1964) and Eglitis (2002).  In all these cases spectral
classification has been done from low dispersion objective prism spectra
and is of low accuracy.  \vskip0.5mm

In the case when one or two ultraviolet color indices of a star were
missing, we have used the same method, but the number of the
$Q$-parameters was smaller:  in the absence of $U$--$V$, $n$ was 10, in
the absence of $U$--$V$ and $P$--$V$, $n$ was 7. Naturally, the accuracy
of classification in these cases for B--A--F--G stars was lower,
especially in luminosity classes.  For two-dimensional classification of
K and M stars the ultraviolet color indices are not essential.
\vskip0.5mm

For the estimation of the classification accuracy in the range of
spectral classes B0--K0, we have made the following test.  The real
program stars were replaced by the test stars having the mean intrinsic
color indices instead of the observed ones.  After that the matching
classification method was applied, taking $\sigma Q$ values $\leq$0.02
mag.

The conclusion is made that spectral classes of B--A--F stars are
determinable with acceptable accuracy when all color indices are
available or only color indices $U$--$V$ are missing.  In the case if
$U$--$V$ and $P$--$V$ color indices are missing, the accuracy of
spectral classes of these stars is very different in various spectral
class ranges. We shall return to a discussion of this accuracy later.

The luminosity classification of B, A and F stars of luminosity classes
V--IV--III is ambiguous even at the presence of all color indices.  The
reason is a lack of one-to-one dependence between MK spectral types and
intrinsic color indices.  Even the unreddened stars of the same
spectral and luminosity classes always exhibit so-called ``cosmic
dispersion" of their color indices and spectral line intensities.  As a
result, both MK and photometric classification of B--A--F stars of
luminosities V--III always have an ambiguity of the order of $\pm$1 of
spectral subclass and $\pm$1 of luminosity class.  When both ultraviolet
color indices are not available, the luminosity classification of
B--A--F--early G stars is impossible.  In this case for late F and early
G stars we accepted that all of them are of luminosity V, which
statistically is not far from reality (F and early G giants of
Population I fall into the Hertzsprung gap in the HR diagram).  These
stars were used in the investigation of interstellar extinction, paying
attention to the lower accuracy of their luminosity classes.
\vskip0.5mm

\WFigure{3}{\psfig{figure=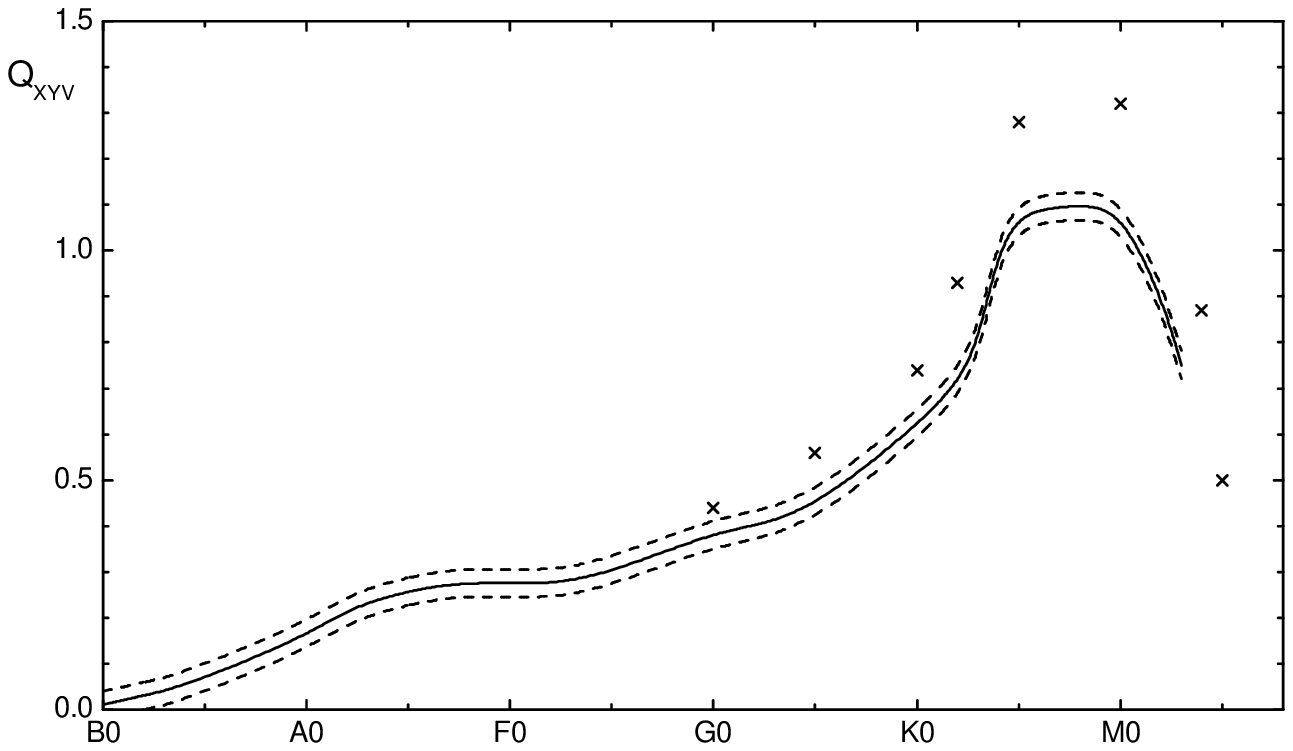,width=10truecm,angle=0,clip=}}
{ The dependence of the interstellar reddening-free
parameter {\itl Q}$_{XYV}$ on spectral class for luminosity V stars.
Crosses are for luminosity III stars.}

In the absence of $U$--$V$ and $P$--$V$ color indices, the best
criterion of spectral class is the $Q_{XYV}$ parameter.  Its dependence
on spectral class is shown in Figure 3. The width of the area between
the two broken lines corresponds to the observed ``cosmic scatter" of
the parameter.  In the B8--A3, F6--G0 and especially in G5--K5 spectral
ranges the $Q_{XYV}$ parameter shows a gradient which is sufficient for
the classification of stars with an accuracy of 1--2 spectral
subclasses.  However, in the range of A5--F5 classes the classification
is of very low accuracy.

\section{5. COLOR EXCESSES, EXTINCTIONS AND DISTANCES OF THE STARS}

     Color excesses $E_{Y-V}$, extinctions $A_V$ and distances $r$ of
the stars were calculated by the equations:
$$
E_{Y-V} = (Y-V)_{\rm {obs}} - (Y-V)_0, \eqno(4)
$$
$$
A_V = R_{YV} E_{Y-V},  \eqno(5)
$$
$$
5\log r = V - M_V + 5 - A_V, \eqno(6)
$$
where the intrinsic color indices ($Y$--$V$)$_0$ for different spectral
and luminosity classes were taken from Strai\v zys (1992, Tables 66-69).
The coefficient $R_{YV}$ for the normal interstellar extinction law is
4.16.  Absolute magnitudes $M_V$ were taken from Strai\v zys (1992,
Appendix 1), according to their spectral and luminosity classes, with a
correction of --0.1 mag, adjusting the old $M_V$ scale to the new
distance modulus of the Hyades ($V$--$M_V$ = 3.3, Perryman et al.
1998). The extinction values and distances of the stars are available in
electronic form from the authors.

The following values of standard deviations $\sigma$ for the determined
quantities are expected (including the cosmic dispersion):
$\pm$0.03--0.04 mag for color excesses, $\pm$0.15 mag for extinctions
and $\pm$25\% for distances.

\section{6. DISTANCES TO THE DUST CLOUDS}

We have divided the investigated area into two parts:  one part is
relatively transparent and includes a part of the North America Nebula,
another one embraces the dark cloud L935.  For both areas the diagrams
$A_V$ versus $r$ are shown in Figures 4, 5 and 6. Figures 4 and 5 are
for the same area, but for different limiting distances. The stars with
lower classification accuracy and the late F -- early G stars classified
without the ultraviolet color indices are shown as crosses.  On all
figures they are scattered in the same area, as the stars with all color
indices available or with only $U$--$V$ missing.  Additionally, in the
diagrams we plotted 105 brighter stars of the same area investigated in
Papers I and III on the ground of their photoelectric photometry.  Their
distances determined in Papers I and III are multiplied by 1.05 to place
them on the same scale.  As a result, total number of stars plotted in
Figures 4/5 and 6 are 354 and 242, respectively.

The dotted curves on the figures show the limiting magnitude effect for
the stars of spectral classes B0\ts V, B2\ts V, B5\ts V, A0\ts V, A5\ts
V, F0\ts V and F5\ts V. The stars of these spectral types above the
corresponding curves are outside accessibility in the present program.
\vskip0.5mm

\WFigure{4}{\psfig{figure=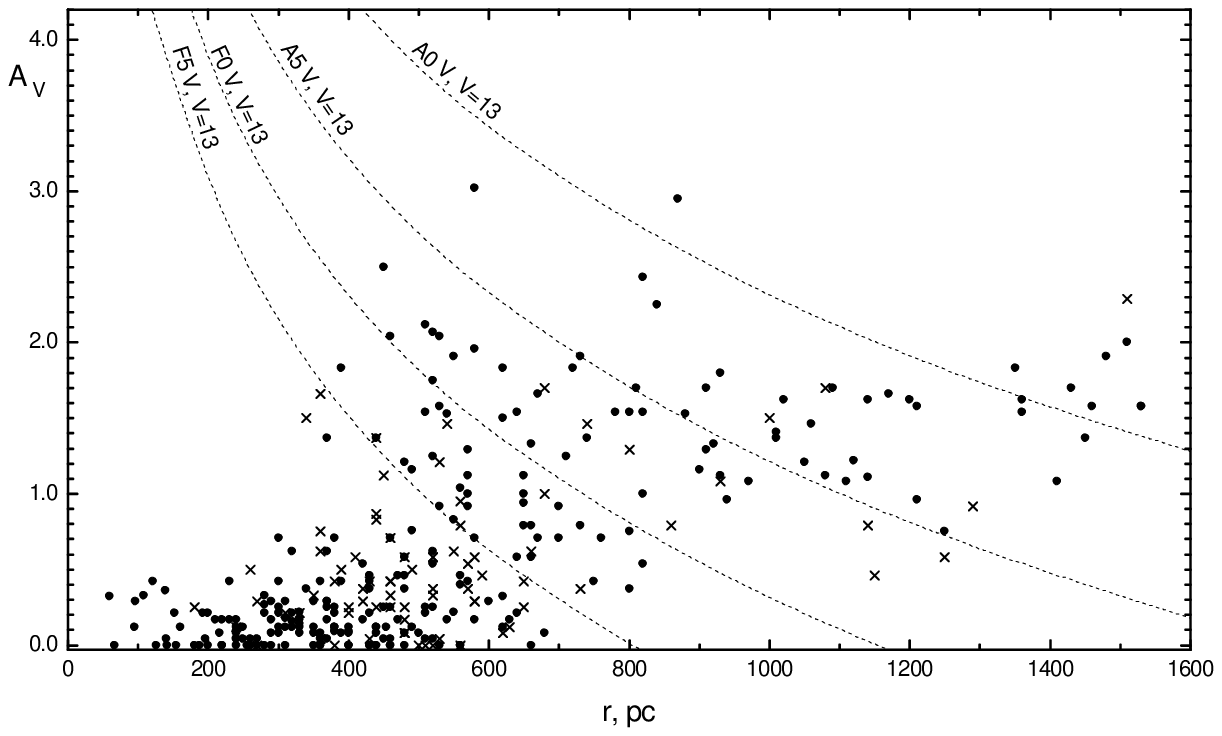,height=7.5truecm,angle=0,clip=}}
{ The dependence of interstellar extinction {\itl A}$_V$
on distance up to 1.6 kpc for the North America Nebula area.}

\WFigure{5}{\psfig{figure=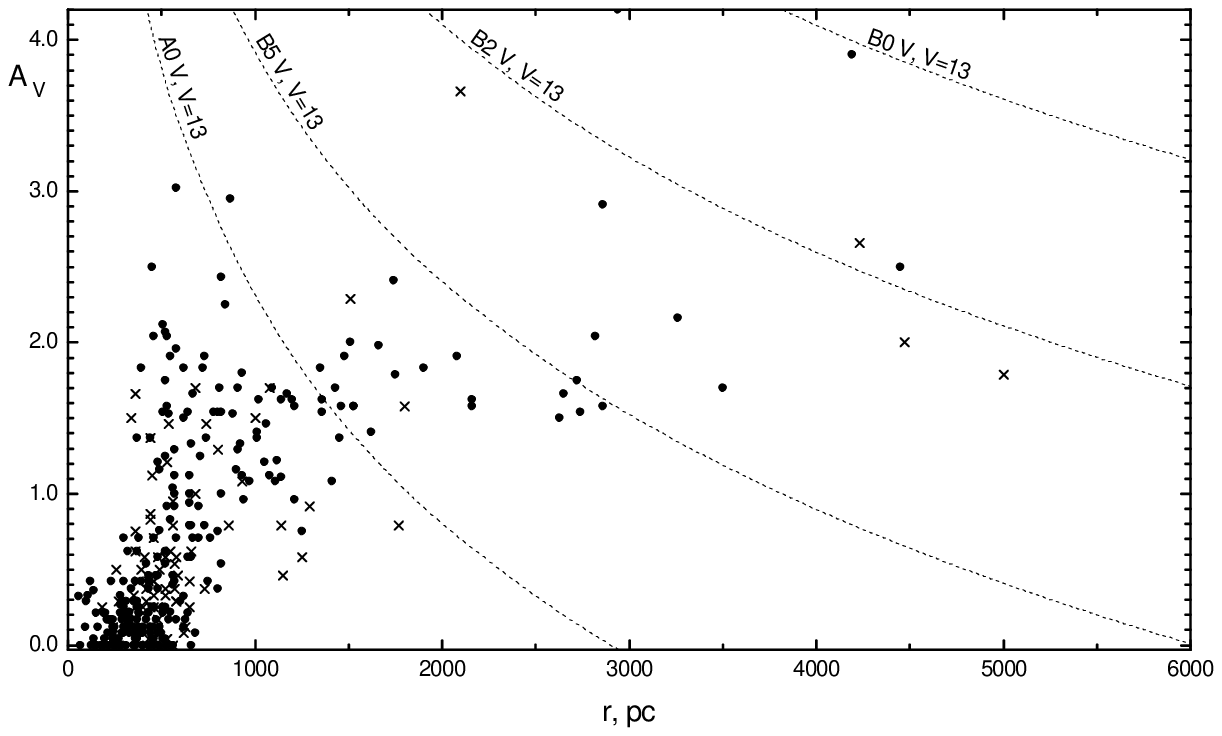,height=7.5truecm,angle=0,clip=}}
{ The dependence of interstellar extinction {\itl A}$_V$
on distance up to 6 kpc for the North America Nebula area.}

\WFigure{6}{\psfig{figure=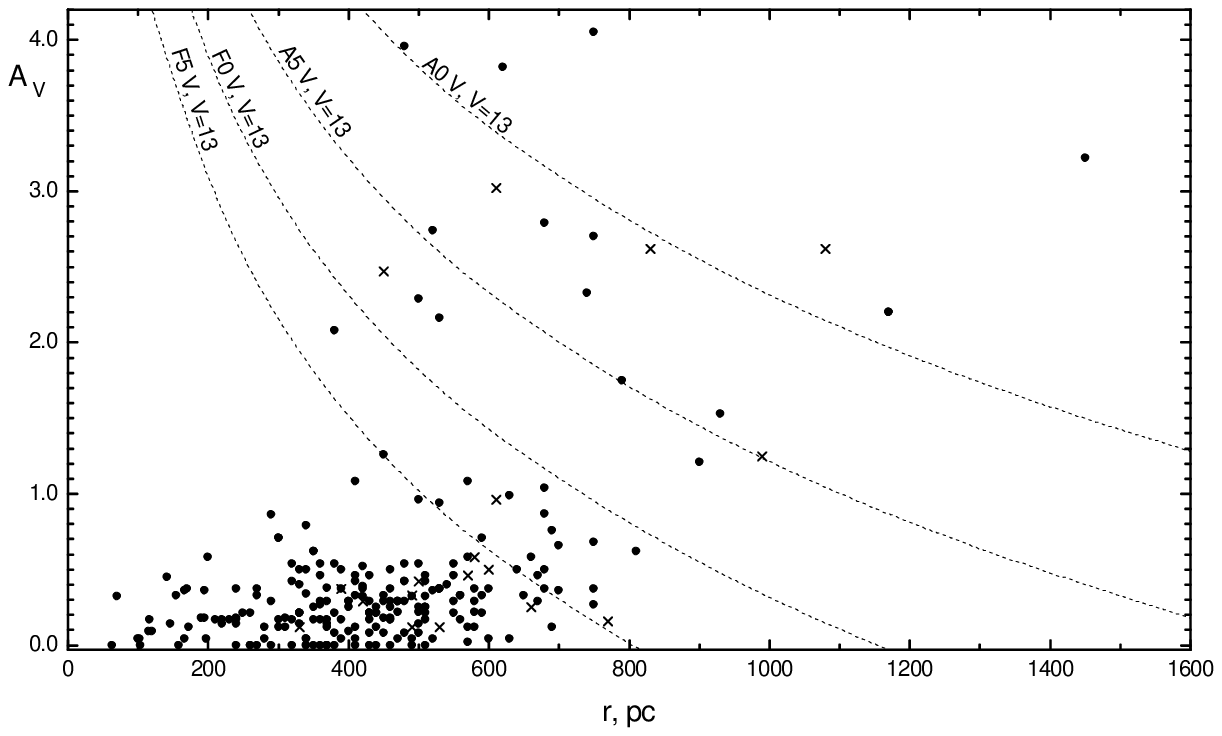,height=7.5truecm,angle=0,clip=}}
{ The dependence of interstellar extinction {\itl A}$_V$
on distance for the area of the dark cloud L935.}

Let us discuss the distribution of stars in the $A_V$ vs.  $r$
for the North America Nebula region shown in Figures 4 and 5. The
following features can be noticed:

(1) The stars with zero reddening are met  from 100 to 600--700 pc. The
same is true for the stars with small reddening, up to $A_V$ = 0.5 mag.

(2) The upper limit of reddened stars gradually increases with
increasing distance, reaching $A_V\sim$0.8 mag at 400 pc.

(3) Approximately at this distance stars with higher extinction start to
appear. The upper limit of $A_V$ is $\sim$4 mag.

(4) The area contains 20 OB-type stars and supergiants at distances
between 2.0 and 6.5 kpc, their $A_V$ values are between 1 and 4 mag.
Most of these stars, if not all, should be inside the Orion spiral
arm.  Our line of view in this direction leaves this arm at about 4 kpc
distance.

Such distribution of stars is consistent with the following model of
distribution of interstellar dust.  Up to a distance of $\sim$600 pc we
see a general Galactic dust layer with an extinction gradient of
$\sim$1.0 mag/kpc.  Approximately at this distance a sharp increase of
dust density takes place.  If we accept a distance error of stars
$\pm$25\%, at 600 pc it corresponds to $\pm$150 pc.  Due to this error,
the expected scatter of stars reddened by the cloud should be observed
between 450 and 750 pc.  This is not far from reality, since we find the
nearest stars with $A_V$$>$1.0 mag at $\sim$400 pc.  \vskip0.5mm

A more accurate distance of the dust cloud may be estimated by taking
the average of limiting distances to the nearest reddened stars and to
the most distant unreddened stars.  We suppose, both these limiting
distances are caused by the same source -- a dust cloud and the distance
errors of stars inside the cloud and just behind it.  The nearest stars
with $A_V$$>$1.0 mag are at 400 pc and the farthest stars with
$A_V$$<$0.5 mag are at 800 pc.  The average of these distances is 600
pc.  Thus, our results are consistent with the distance of the absorbing
dust cloud at 600 pc, in good accordance with the cloud distance
determined in Paper III.  \vskip0.5mm

Maximum extinction in the direction of the North America Nebula is not
high:  a lot of faint stars are seen on deep photos.  Among the stars
with the extinction $A_V>1.5$ mag, 31 are closer than 1 kpc, 22 are
between 1 and 2 kpc and 20 are farther than 2 kpc.  Stars which are
closer than 1 kpc exhibit the maximum extinction values at $\sim$3 mag.
\vskip0.5mm

Now let us turn to the $A_V$ vs.  $r$ graph for stars in the dark
cloud area, shown in Figure 6. Although both transparent and dark areas
are of comparable apparent size, the last one shows much smaller
surface density of stars, and most of them are the foreground objects.
Actually both areas are very similar with respect to the number of
foreground stars and their distribution on the $A_V$ vs.  $r$ diagrams.
However, the dark area is very poor of stars with extinctions larger
than 1.0 mag at distances farther than 800 pc.  In the dark area, the
nearest reddened stars with $A_V>1.0$ mag and the farthest stars with
$A_V<0.4$ mag are almost at same distances as in the transparent area
discussed above:  at 400 and 800 pc.  Their average distance is 600 pc.
It is difficult to estimate the error of the cloud distance since we do
not know which stars reside in the cloud and which are behind it.
Probably the distance is accurate within $\pm$50 pc.
\vskip0.5mm

Thus, we are safe to accept that the absorbing clouds in both areas are
at the same distance.  However, in the dark area the extinction is much
larger, and here down to 13 mag we see only a few background stars.  The
majority of reddened stars with $A_V>1.5$ mag are situated within 400
and 800 pc, the limiting distances for stars residing within the dust
cloud.  The scarce background stars probably are seen through
semitransparent cloud windows.  In other directions the cloud in the
optical wavelengths is almost black.  According to Cambr\'esy et al.
(2002), in some directions of our area the visual extinction may be
considerably larger than 5 mag.  \vskip0.5mm

We have no idea how deep is the L935 dust cloud, i.e., is there a single
comparatively thin dust sheet or the cloud has extensions along the line
of sight.  If the cloud is approximately round, its thickness should not
exceed 20--30 pc.

\section{7. RESULTS AND CONCLUSIONS}

(1) CCD photometry in the {\it Vilnius} seven-color system has been done
for 690 stars down to 13.2 mag in the 2$\times$2 sq. degree area
including the North America Nebula and the dust cloud L935, separating
the North America and Pelican nebulae. About 150 of these stars have
been observed photoelectrically.

(2) Majority of the stars have been classified in spectral and
luminosity classes. Their color excesses, interstellar extinctions and
distances have been determined.

(3) The extinction vs. distance graphs have been plotted separately for
the area of the North America Nebula and for the dust cloud L935 area.
It is shown that the interstellar extinction in both areas up to 600 pc
distance is consistent with the general Galactic dust layer with a
gradient of $\sim$1.0 mag/kpc.

(4) A steep increase of extinction is observed in both areas at
$\sim$400 pc, which may be explained by the presence of a dust cloud at
about 600 pc distance.  In the area of the North America Nebula this
cloud is relatively thin, its extinction does not exceed $A_V$ = 3 mag.
In the area of the dark cloud L935 the extinction is much larger.  Due
to the limiting magnitude effect, the stars with extinctions greater
than 4 mag are not observable in both areas.

(5) This work shows that the Maksutov telescope of the Mol\.etai
Observatory is an excellent instrument for a precise CCD photometry
since in one exposure it gives a 1.2$\times$1.2 sq. degree field with a
25$\times$25 sq. mm CCD chip.  However, the meniscus lens of the
telescope should be replaced to one, more transparent in the
ultraviolet. Also, a CCD chip with enhanced ultraviolet sensitivity is
to be used.

\vskip5mm

ACKNOWLEDGMENTS.  We are grateful to Ilgmars Eglitis (Baldone
Observatory, Latvia) for classification of some stars from
objective-prism plates, A. G. Davis Philip for reading the manuscript
and for valuable corrections and to Justas Zdanavi\v cius for assistance
in observations.  We acknowledge the use of the SkyView facility located
at NASA Goddard Space Flight Center and the Simbad database of the
Strasbourg Stellar Data Center.

\References

\ref Cambr\'esy L., Beichman C.\ts A., Jarrett T.\ts H., Cutri R.\ts M.
2002, AJ, 123, 2559

\ref Eglitis I. 2002, personal communication

\ref Kharadse E.\ts K., Apriamashvili S.\ts P., Kochlashvili T.\ts A.
1964, Bull. Abastumani Obs., No. 31, 3

\ref Metik L.\ts P. 1960, Izvestia Crimean Obs., 23, 60

\ref Perryman M.\ts A.\ts C., Brown A.\ts G.\ts A., Lebreton Y. et al.
1998, A\&A, 331, 81

\ref Schwassmann A., van Rhijn P.\ts J. 1938, Bergedorfer
Spektral-Durch\-mus\-te\-rung der 115 N\"ordlichen Kapteynschen
Eichfelder, Vol. 2, Hamburger Sternwarte in Bergedorf, p. 293 (SA 40)

\ref Strai\v zys V. 1983, Bull. Vilnius Obs., No. 62, 11

\ref Strai\v zys V. 1992, {\itl Multicolor Stellar Photometry}, Pachart
Publishing House, Tucson, Arizona

\ref Strai\v zys V., Corbally C.\ts J., Laugalys V. 1999, Baltic
Astronomy, 8, 355

\ref Strai\v zys V., Goldberg E.\ts P., Mei\v stas E., Vansevi\v cius V.
1989b, A\&A, 222, 82 (Paper II)

\ref Strai\v zys V., Kazlauskas A. 1993, Baltic Astronomy, 2, 1

\ref Strai\v zys V., Kazlauskas A., Boyle R.\ts P., Vrba F.\ts J.,
Smriglio F. 1996, Baltic Astronomy, 5, 165

\ref Strai\v zys V., Kazlauskas A., Vansevi\v cius V., \v Cernis K.
1993, Baltic \hfil\break Astronomy, 2, 171 (Paper III)

\ref Strai\v zys V., Mei\v stas E., Vansevi\v cius V., Goldberg E.\ts P.
1989b, Bull. \hfil\break Vilnius Obs., No. 83, 3 (Paper I)

\ref Strai\v zys V., Sviderskien\.e Z. 1972, Bull. Vilnius Obs., No. 35,
3

\bye